\documentclass[CEJP,PDF]{cej} 
\usepackage{layout}
\usepackage{amsmath}
\usepackage{textcomp}
\usepackage{hyperref}
\usepackage{type1cm}

\title{Application of High Intensity THz Pulses for Gas High Harmonic Generation}

\articletype{Research Article}

\author{Emeric~Balogh\inst{1}\email{ebalogh@titan.physx.u-szeged.hu},
        J\'ozsef A.~F\"ul\"op\inst{2},
        J\'anos~Hebling\inst{2,3},
        P\'eter~Dombi\inst{4},
        Gy\H{o}z\H{o}~Farkas\inst{4},
        Katalin~Varj\'u\inst{1}\email{varju@physx.u-szeged.hu}
       }

\institute{
     \inst{1} Department of Optics and Quantum Electronics, University of Szeged,\\
     Dom t\'er 9, 6720 Szeged, Hungary
     \inst{2} MTA-PTE High-Field THz Research Group,\\
     Ifj\'us\'ag u. 6, 7624  P\'ecs, Hungary
     \inst{3} Department of Experimental Physics, University of P\'ecs,\\
     Ifj\'us\'ag u. 6, 7624  P\'ecs, Hungary
     \inst{4} Wigner Research Centre for Physics,\\
     Konkoly-Thege M. \'ut 29-33, 1121 Budapest, Hungary
     }

\abstract{The main effects of an intense THz pulse on gas high harmonic generation are studied via trajectory analysis on the single atom level. Spectral and temporal modifications to the generated radiation are highlighted.}


\begin{document}
\maketitle


\section{Introduction}

Since the production of attosecond pulses was experimentally verified in 2001 \cite{r1,r2}, a great deal of effort has been devoted to fully understand and improve high harmonic generation in gases to provide a versatile source of coherent XUV to X-ray radiation \cite{r3}. 
The process relies on optically ionizing rare gases in intense laser fields then steering the ionized electrons with the laser field. 
The freed electron spends about half a period in the laser field and under special circumstances it has a chance to recombine with the ionic core, leading to the emission of a high energy photon. 
A wide spectrum of radiation with approximately constant amplitude is produced. This spectrum ends in a cutoff whose position is defined by the properties of the laser pulse and that of the ionized gas. 
For a multicycle laser pulse, the repetition of the process leads to the appearance of odd harmonics in the spectrum whereas in the time domain, a train of attosecond pulses are observed. 
In very special circumstances, harmonic generation happens only once during the laser pulse and an isolated attosecond burst is produced.
A limitation of the use of this radiation is the rather low efficiency of harmonic production (typically in the range of 10$^{-5}$ - 10$^{-6}$). 

Source improvement efforts are as follows: (i) to increase bandwidth, reaching higher frequencies and also effecting the synthesized attosecond pulse duration; (ii) to increase the efficiency of the process and (iii) to reduce the number of pulses or their repetition period. 
In this paper, we report our studies on the effect of a high intensity THz pulse on high harmonic generation by a short and intense laser pulse.

\section{Method}

To understand the process of high order harmonic generation, we use the non-adiabatic saddle point method \cite{r4}. 
This model calculates the electric dipole driven by the laser field using the one electron, one bound state approximation while treating the strong laser field classically:
\begin{align}
x(t)=i \int_{-\infty}^{t}dt' \int d^3p\:E(t') d^*[p-A(t)] d[p-A(t')] exp[-iS(p,t,t')]+c.c.
\end{align}
where E(t) is the laser electric field, A(t) is the vector potential, p is the electron's canonical momentum and S the semiclassical action defined as:
\begin{align}
S(p,t,t')=\int_{t'}^{t}dt'' \: \Bigg(\frac{[p-A(t'')]^2}{2}+I_p \Bigg)
\end{align}
Trajectories with a stationary phase are determined
\begin{align}
\begin{array}{l l l}
    p_s=\frac{1}{t-t'}\int_{t'}^t A(t'') dt'' & \quad \frac{[p_s-A(t'_s)]^2}{2}+I_p=0 & \quad \omega-\frac{[p_s-A(t_s)]^2}{2}-I_p=0
\end{array}
\end{align}
where t$'$ is the ionization time and t is the return time of the electron. 
These equations resemble energy and momentum conservation for the ionization, forced oscillation and recombination of the electron \cite{r5}.

Analyzing the electron trajectories yields information on spectral and temporal characteristics of the radiation. 
The characteristic parameters (ionization time, recombination time and kinetic energy at the moment of recombination) of the driven electrons provide in-depth information on the high-harmonic radiation. This is illustrated in Fig. 1. for an 800 nm laser pulse of intensity 6*10$^{14}$ W/cm$^2$. 
With inversion symmetry, the half-cycle periodicity of the laser electric field leads to a half-cycle periodicity of the produced radiation. 

Electrons with different kinetic energy return at different times leading to a delay of the spectral components. 
In each half-cycle, a class of short trajectories and a class of long trajectories produce radiation up to the cutoff frequency \cite{r6}. 
Macroscopic separation of the radiation yields positively and negatively chirped pulses differently delayed with respect to the electric field. 
The harmonic bursts produced in consecutive half-cycles are equivalent due to the inversion symmetry of the process.

\begin{figure}
\includegraphics[width=14cm]{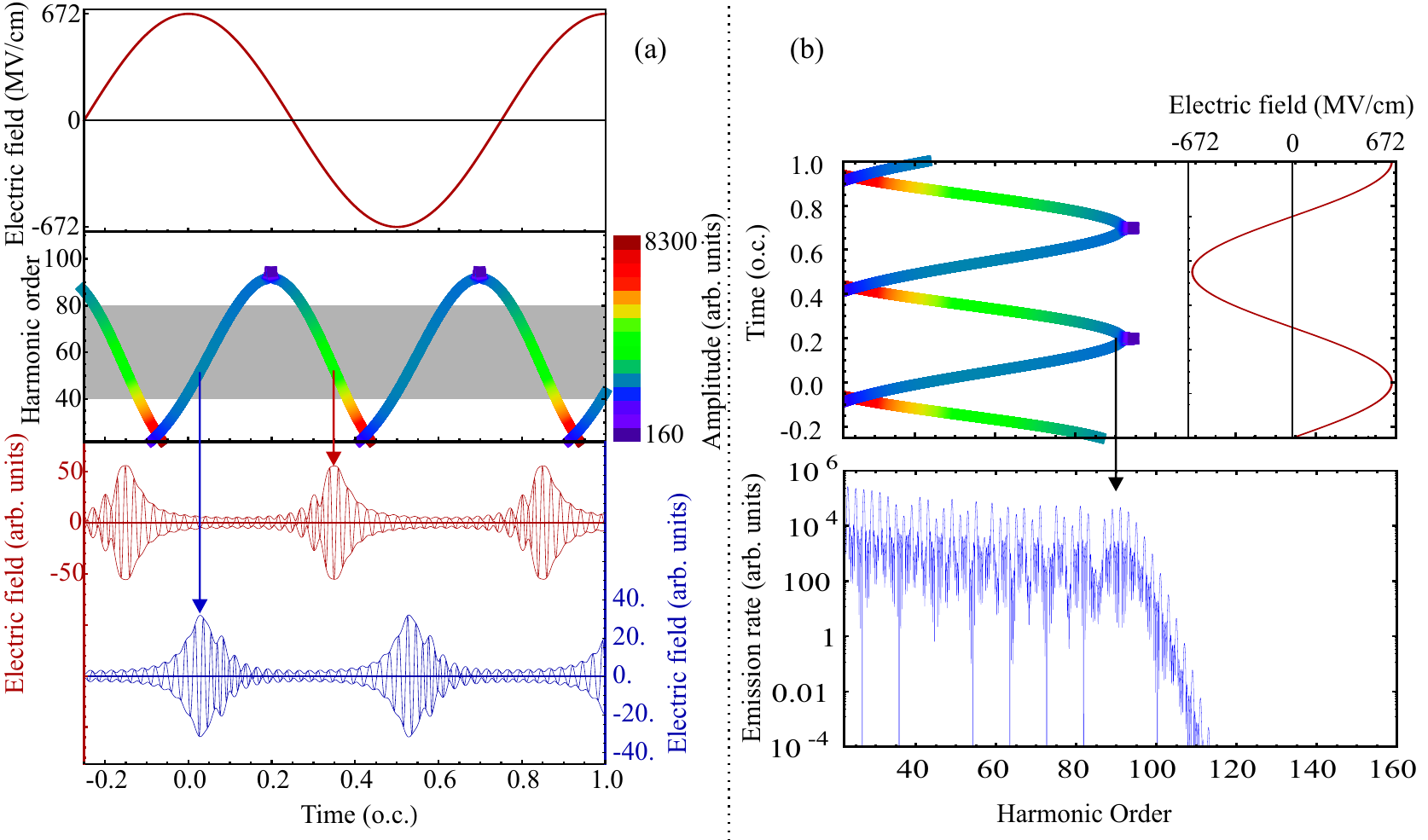}
\caption{Temporal (a) and spectral (b) information obtained from trajectory analysis. (a) Electron trajectories (middle) and attosecond pulses (bottom) illustrated in time with respect to the laser electric field (top). The color scale shows the dipole strength corresponding to trajectories with the indicated return time and energy (expressed in photon number). Short (long) trajectories produce the positively (negatively) chirped blue (red) pulses. The gray shaded area shows the spectral region used for reconstruction of attosecond pulses. (b) The cutoff where the short and long trajectories meet corresponds to the sharp end of the spectrum.}
\end{figure}

\section{Effect of the THz field}

With a recent major advance in THz pulse generation via difference frequency generation in GaSe \cite{r7}, electric field strengths up to 100 MV/cm at around 30 THz have previously been achieved. 
These field strengths reach up to ten percent of the laser field strength, thus their application for high harmonic generation can significantly modify electron trajectories and hence the generated radiation. 
The appearance of these long wavelength pulses provides a means to achieve an alteration of the HHG process which was previously suggested using a static electric field \cite{r8,r9,r10,r10a,r11}.

\begin{figure}
\includegraphics[width=14cm]{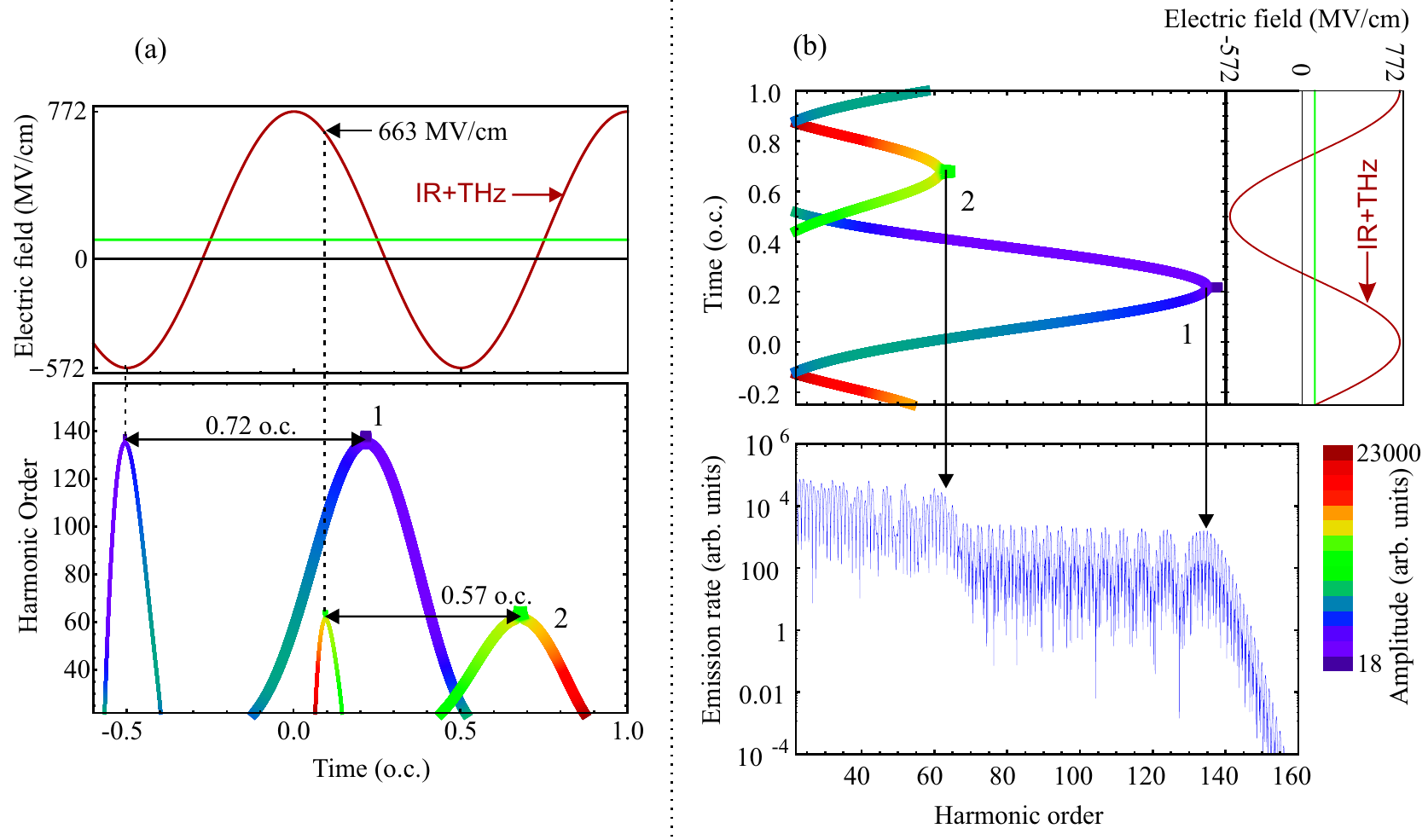}
\caption{Electron trajectories and high harmonic spectrum generated in the presence of the THz field. THz field strength is 100 MV/cm indicated in green, the 6*10$^{14}$ W/cm$^2$ laser intensity corresponds to 672 MV/cm. Breaking the symmetry leads to alternating properties of the radiation produced. In (a), ionization (thin lines) and recombination (thick lines) times of electron trajectories from a single optical cycle (o.c.) are presented to show the generation of the double plateau structure. Electron trajectories noted with (1) start at a weaker field strength ($\approx$570 MV/cm), and travel longer before recombination, hence the lower emission rate. However the strong field present in the next half-cycle allows them to gain more kinetic energy before recombination, producing a higher cutoff. On the other hand electrons released in the next half-cycle (2) start at much higher field strengths ($\approx$ 660 MV/cm), and recombine sooner resulting in a higher emission rate but the cutoff is reduced as the THz field prevents them gaining kinetic energy. (b) The two cutoffs with different emission rates produce the double-plateau structure of the spectrum.}
\end{figure}

The first and obvious effect of the presence of the THz field is breaking of the inversion symmetry leading to a full cycle periodicity and the appearance of both odd and even harmonics (Fig. 2). 
Our trajectory analysis provides full understanding of how the trajectories are modified in the consecutive half-cycles. 
In one half-cycle we observe trajectories corresponding to weaker dipole strengths and ending in a higher cutoff (1), in the other half-cycle, the trajectories meet at a lower cutoff but the corresponding dipole strengths are stronger (2).
The difference in dipole strengths is a consequence of the direct relationship between ionization probability and the instantaneous electric field strength at the time of ionization. Diffusion effects also play a role as reported in \cite{r9,r10a}.
On the other hand, the cutoff varies with the amplitude of the electric field during the free travel of the electron in the next half cycle.
A time resolved spectrometer should see a half-cycle variation of the spectrum, strong with narrower and weak with broader bandwidth. 
Using a time-integrating spectrometer a double-plateau spectrum is recorded, where the two cutoffs are symmetric about the THz-free cutoff (Fig.3).

\begin{figure}
\includegraphics[width=6.5cm]{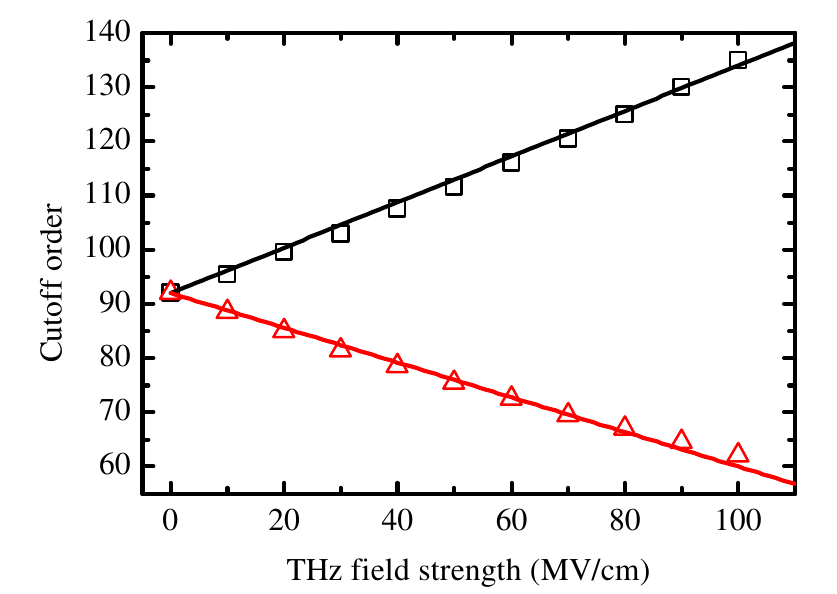}
\caption{Variation of the higher (black squares) and lower (red triangles) cutoffs with THz field strength. Higher cutoffs are generated when the THz field, having the same sign as the IR, increases the kinetic energy of the recombining electrons.}
\end{figure}

If the radiation is spectrally filtered from the lower plateau, a harmonic pulse is produced every half-cycle with strongly alternating amplitude, whereas from the higher plateau only one pulse per cycle is synthesized.

\section{Applications of an assisting THz field to improve a high harmonic source}

In the previous section we illustrated the changes a THz pulse imposes on the harmonic radiation. 
We can conclude that the effects are manifold. Spectrally we observe an increase in amplitude in the lower plateau and a large extension of the cutoff. 
In the time domain the repetition of the attosecond bursts are modified. 
A comparison of the trajectories also shows that in the usual, laser only case, the long trajectories carry the larger portion of radiation, whereas in the THz assisted case, there is a redistribution of electrons to the shorter trajectory class.

\begin{figure}
\includegraphics[width=14cm]{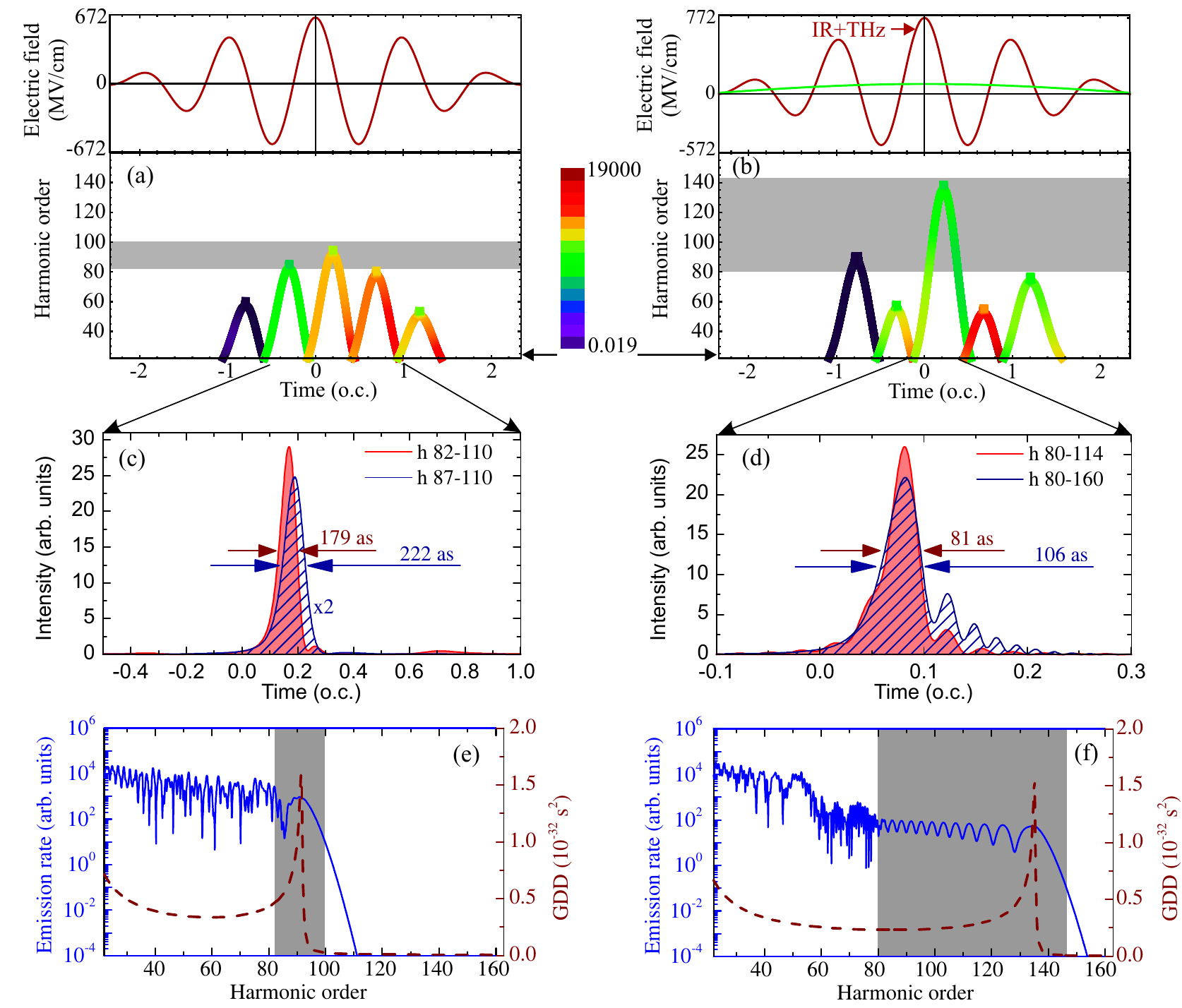}
\caption{Trajectories (a,b), attosecond pulses (c,d) and spectra (e,f) produced by a few-cycle (5.2 fs) laser pulse in the absence (left column) and presence (right column) of the THz pulse. The spectral range indicated by gray shading can be used for single attosecond pulse synthesis. In (c) and (d) attosecond pulses are shown, synthesized from harmonics indicated in the legends. In the absence of the THz pulse (c), the shortest SAP of 179 as is produced from harmonics above 82 (red, shaded area), when the presence of a weak second pulse (generated in the next half-cycle) is also allowed. Because of the inherent group delay dispersion (GDD) of high-order harmonics in the presence of the THz pulse, the shortest SAP of 81 as is generated from a narrower spectral region (harmonics 80 to 114) than that which is available. Note the different time-scales in (c) and (d).
Dark red, dashed lines on (e) and (f) show the GDD of short trajectory components from the middle half-cycle of the laser pulse. Long trajectories have similar GDD values (not shown) but with opposite sign.}
\end{figure}

A very important effect for improving the attosecond source is realized when the THz field is superimposed on a short, few-cycle laser pulse. 
Using a now routinely available 5.2 fs pulse, the trajectories are modified such that the highest harmonics are produced only in a single halfcycle and a very broad continuum can be spectrally filtered to produce isolated attosecond pulses. 
However, as the high-harmonic radiation possesses an intrinsic chirp (see Fig. 4) due to the individual emission times of different harmonics, transform limited pulses are only achievable using post-compression methods or when using only cutoff harmonics for synthesis.

The shortest isolated attosecond pulses obtainable (using only short and/or cutoff trajectory radiation) in the laser only case presented in Fig 4, is 222 as (218 as transform limit) synthesized from harmonics 87 to cutoff ($\approx$97). 
If we allow an SAP with a slightly worse contrast ratio of 1:50 (i.e. a small second pulse is also present), the SAP duration can be reduced to 179 as (with 159 as transform limit), synthesized from radiation above harmonic 82 (see Fig. 4.c). 
In the THz assisted case, the shortest SAP without compression is generated from harmonics 80 to 114 resulting in an 81 as long SAP with a transform limited duration of 73 as (see Fig. 4.d). 
However, in this case a much wider spectrum can be used for SAP production if post-compression methods are available, reducing the shortest theoretically achievable SAP duration to 46 as.

Increasing the bandwidth for single attosecond pulse production might be the most important effect of combining the strong THz field with the laser pulse.

\section{Discussion}

The work presented above illustrates the potential in using THz pulses to assist high harmonic generation by laser pulses. 
These calculations were carried out using trajectory analysis on the single atom level.
For a true characterization of the radiation produced in a macroscopic medium, propagation of the generating and harmonic fields needs to be included in the model \cite{r12}. 
Full 4D calculations are more realistic in their results, however yield less insight in the physical process that we attempted to illustrate above. 

\section*{Acknowledgments}

The project was supported by the European Community's Seventh Framework Programme under contract ITN-2008-238362 (ATTOFEL) and by the Hungarian Scientific Research Fund (OTKA) grant number NN 107235.
KV and PD acknowledge the support from the Bolyai Foundation. 
KV is also grateful for the support of NKTH-OTKA (\#74250).

\end{document}